\documentclass[a4paper,11pt]{article}
\pdfoutput=1 % if your are submitting a pdflatex (i.e. if you have
\usepackage{jcappub} % for details on the use of the package, please
\usepackage[T1]{fontenc} % if needed

\title{\boldmath Lunar system constraints on the modified theories of gravity}

\author{Qasem Exirifard}

\affiliation{
Institute for Research in Fundamental Sciences (IPM),
\\Tehran, Iran
}

\emailAdd{exir@theory.ipm.ac.ir}

\abstract{The MOND paradigm to the missing mass problem requires introducing a functional that is to be identified through observations and experiments. We consider the AQUAL theory as a realization of the MOND.   We show that the accurate value of the Earth GM measured by  the Lunar Laser Ranging measurements   and that  by various artificial Earth satellites,  including the accurate tracking of the LAGEOS satellites,  constrain this functional such that some of the chosen/proposed functional are refuted. }

\begin{document}
\maketitle
\flushbottom

\section{Introduction}
The missing mass problem in galaxies can be resolved either by the Modified  Newtonian Dynamics (MOND), or the Modified Gravity  (MOG) or the dark matter hypothesis.  The first two paradigms require introducing a functional that interpolates the Newtonian regime to the MOND regime. This functional is to be identified by examining the data.

%\section{Lunar System constraints }
We note that the accurate value of the mass ratio of the Sun/(Earth+Moon) from the Lunar Laser Ranging combined with the Solar GM and the lunar GM from lunar orbiting spacecrafts \cite{LLR} yields the effective   gravitational mass of the Earth in an Earth-centered reference frame with the precision of one part in $10^8$ :
\begin{equation}\label{GMLLR}
GM^{LLR}_{Earth}(r_{_{LD}}) = 398600.443\pm 0.004
\frac{km^3}{s^2}
\end{equation}
where $r_{_{LD}}$ represents the Lunar distance: the average distance  between the Earth and Moon. The effective gravitational mass is defined to be the gravitational field multiplied by $r^2$ where $r$ is distance from the center of the earth.  The effective gravitational mass of the Earth has also been measured by various artificial Earth satellites , including the accurate tracking of the LAGEOS
satellites orbiting the Earth in nearly circular orbits with
semimajor axes about twice the radius of the Earth \cite{LAGEOS}:
\begin{equation}\label{GMLAGEOS}
GM^{LAGEOS}_{Earth}(2 r_{Earth})=  398600.4419 \pm 0.0002 \frac{km^3}{s^2}
\end{equation}
where $r_{Earth}$ stands for the radius of the Earth.

 In this paper, we study the earth Lunar system constraints on the interpolation function. In section \ref{MOND}, we first briefly review the MOND paradigm.  We present the choices for the interpolating function. The section \ref{MONDLaser} proves that a relativistic version of MOND does not affect the null geodesics at the precision of the Lunar Laser Ranging measurements.  The leading effect of MOND is changing the ratio of the `Earth-Moon distance' to the `LAGEOS satellites-Earth distance'.  In section \ref{constraints}, we translate the consistency of  \eqref{GMLLR} and \eqref{GMLAGEOS} to the first Lunar system constraint on the MOND. We show that the Lunar system constraints refute some of the choices of the interpolating function.

\section{A Brief Review of  MOND}
\label{MOND}
MOND \cite{MOND} alters the Newton's second law of dynamics to
\begin{equation}
F\,=\, m f(\frac{|a|}{a_0}) \vec{a}\,,
\end{equation}
wherein $F$ is the total force exerted on the particle and  \cite{MONDa0}
\begin{equation}\label{a0}
a_0 = 1.0  \times 10^{-10} \frac{m}{s^2}\,.
\end{equation}
and $f(x)$ is a function possessing  the following asymptotical behaviors 
\begin{eqnarray}
f(x) = \left\{
\begin{array}{ccc}
1&,& x\gg 1 \\
x &,& x \leq 1
\end{array}
\right.\,.
\end{eqnarray}
 MOND is to be applied in a frame inertial with respect to the frame wherein the CMB background is isotropic. 
The second law of Newtonian dynamic  has been tested in a lab on the Earth surface down to the acceleration of $10^{-11}\frac{m}{s^2}$ \cite{Abramovici:1986eu}, and $10^{-14}\frac{m}{s^2}$ \cite{Gundlach:2007zz}.  No deviation has been observed. However,  these experiments are not performed in a frame inertial with respect to the CMB frame. They are perfumed in an accelerating frame with respect to the CMB frame: the Earth. So they imply no conclusive constraint on the MOND.

The second approach is theories of modified gravity. We consider the AQUAL (aquadratic Lagrangian theory)  model \cite{AQUAL}. AQUAL alters the Newtonian gravitational potential equation
\begin{equation}
\nabla^i\nabla_i \Phi_N  = 4 \pi G \rho,
\end{equation} 
to
\begin{equation}\label{phiMOND0}
\nabla^i\left( \mu(\frac{|\nabla \Phi|}{a_0}) ~\nabla_i \Phi \right)= 4 \pi G \rho,
\end{equation}
 where $\mu$ is a functional  of the gravitational field strength, and $a_0$ is given by  \eqref{a0}.  The asymptotic behavior of $\mu$ is required to coincide to that of $f$:
\begin{eqnarray}
\label{mux}
\mu(x) = \left\{
\begin{array}{ccc}
1&,& x\gg 1 \\
x &,& x \leq 1
\end{array}
\right.\,.
\end{eqnarray}
Ref.  \cite{Meyer:2011ae} reports that the Newtonian  law of gravitation is true at a gravitational acceleration at order of $a_0$ in Earth. However,  this experiment is performed  in the presence of the gravitational field of the Earth. So it implies no conclusive constraint on the AQUAL model. Notice that  the AQUAL model fixes only the asymptotic behavior of the functional $\mu$. Various functionals possessing these asymptotic behaviors have been suggested for $\mu$. We consider all the families of the suggestions reviewed in \cite{Famaey:2011kh}:
\begin{subequations}\label{literature}
\begin{eqnarray}
\mu_1(x) &=& \frac{(1+4x^2)^{1/2}-1}{2x},
\label{unruhmu}\\
\mu_2(x) &=& 1 - (1+x/3)^{-3}\,,\\
\mu_{\alpha}(x) &=& \frac{2x}{1+ (2-\alpha) x + [(1-\alpha x)^2 + 4x]^{1/2}}\,,
\label{alphafamily}
\\
\mu_n(x)&=&{x\over (1+x^n)^{1/n}}\,,
\label{nfamily}\\
\label{nun}
\nu_n(y)&=&\left[{1+(1+4y^{-n})^{1/2}\over 2}   \right]^{1/n}\,,\\
\label{nubeta}
\nu_{\beta}(y)&=&(1-e^{-y})^{-1/2} + \beta e^{-y}\,,\\
\label{nugamma}
\nu_{\gamma}(y)&=&(1-e^{-y^{\gamma/2}})^{-1/\gamma}+(1-\gamma^{-1})e^{-y^{\gamma/2}}\,,\\
\label{nudelta}
\nu_{\delta} (y) &=& (1-e^{-y^{\delta/2}})^{-1/\delta}\,,
\end{eqnarray}
\end{subequations} 
where 
\begin{equation}
\nu(y)=\frac{1}{\mu(x)} \, ,  
\end{equation}
where
\begin{equation}\label{y=xmux}
 y=x\mu(x).
 \end{equation}
 Note that $\mu_1$, $\mu_2$, $\mu_\alpha$, $\mu_n$ and $\nu_n$ connect the MOND regime to the newtonian regime by a power law interpolating function. $\nu_\beta$, $\nu_\gamma$ and $\nu_\delta$ connect these regimes  by an exponentially suppressed interpolating function. We expect that studying the gravity in the Newtonian regime constrains  $\mu_1$, $\mu_2$, $\mu_\alpha$, $\mu_n$ and $\nu_n$. To constrain $\nu_\beta$, $\nu_\gamma$ and $\nu_\delta$ one should study the MONDian regime of the theory. 
 
\section{MOND Effects on the Lunar Laser Ranging Measurements }
\label{MONDLaser}
The relativistic versions of MOND attribute the MOND effects to the physical metric that particles feel. They assume that the metric is not what is predicted by the Einstein gravity. 
 Let the space-time geometry around the Earth coincides to that predicted by the Einstein gravity:
\begin{equation}
ds^2= -g_{\mu\nu} dx^\mu dx^\nu \,.
\end{equation}
A modified theory of gravity adds a correction to the physical metric near the Earth:
\begin{equation}
ds^2= g_{\mu\nu}^{phy}  dx^\mu dx^\nu = (g_{\mu\nu}+ g_{\mu\nu}^{(m)}) dx^\mu dx^\nu \,,
\end{equation}
where $g_{\mu\nu}^{(m)}$ is the deviation from the prediction of the Einstein theory of gravity. 

The leading metric predicted by the Einstein theory can be approximate to the Schwarszchild metric:
\begin{eqnarray}
g_{\mu\nu} &=& g_{\mu\nu}^{(0)}  + \epsilon g_{\mu\nu}^{(1)} + \cdots\,,\\
g_{\mu\nu}^{(0)} dx^\mu dx^\nu &=&  (1- \frac{r_h}{r})c^2 dt^2 + \frac{dr^2 }{1- \frac{r_h}{r}} + r^2 d\Omega^2\,,\\
r_h & = & \frac{2 G M_\oplus}{c^2}\,,
\end{eqnarray}
where the $\epsilon$ terms are due to inhomogeneity of the Earth, and also encode the contribution of the Sun and other objects.  We call these terms the $\epsilon$ terms. We assume that the $\epsilon$ terms  are sub-leading. In other words the effect of the external gravitational field is sub leading too.  We, further, postulate that 
\begin{eqnarray}\label{epsilonexpansions}
g^{(m)}_{\mu\nu} = g^{(m)(0)}_{\mu\nu} + \epsilon g^{(m)(1)}_{\mu\nu}  + \cdots\,,
\end{eqnarray}
In other words the deviation from the Einstein prediction inherits  the $\epsilon$ expansion series. 

Near the Earth the leading term is the Schwarszchild metric. We truncate the series in $\epsilon$ expansion of the metric to the leading term. This implies the  truncation of the $\epsilon$ expansion of  $g^{(m)}_{\mu\nu}$ to the leading term. The $\epsilon$ expansion in $g_{\mu\nu}^{phy} $ is not needed for the leading approximation.  In this approximation, $g_{\mu\nu}$ possesses spherical symmetry and is stationary. We assume that the corrections due to the considered modified theory of the gravity should respect these symmetries. So within this approximation $g_{\mu\nu}^{(m)}$ should be spherical and stationary:
\begin{eqnarray}
ds^2 &=& (g_{\mu\nu}+ g_{\mu\nu}^{(m)}) dx^\mu dx^\nu  \nonumber \\
 &=&- \left(1- \frac{r_h}{r}+ g_{tt}^{(m)(0)}\right) c^2 dt^2
 +\left( \frac{1}{1- \frac{r_h}{r}} + g_{rr}^{(m)(0)} \right) dr^2
 + r^2 d\Omega^2\, + O(\epsilon)\,.
\end{eqnarray}
 The AQUAL theory only tells us what $g_{tt}^{(m)(0)}$ is. Knowing $g_{tt}^{(m)(0)}$ suffices to address the orbits of non-relativistic objects/probes.  However, to address the Null geodesics it is necessary to know $g_{rr}^{(m)(0)}$.  Since lasers are used to perform accurate tracking of the Moon and the LAGEOS satellites we must be cautious. We, therefore, estimate  the contribution of the modified gravity on measuring long distances by lasers.  We prove that this contribution can be consistently neglected at the precision required in this work.
 
The  general solution to the AQUAL model can be expressed in term of the corresponding Newtonian gravitational potential 
\begin{eqnarray}\label{EMond}
\mu(\frac{|\nabla\Phi|}{a_0}) \nabla\Phi  & = & \nabla\Phi_{N} + \nabla\times \vec{{h}} \,,
\end{eqnarray}
where $\vec{{h}}$ is identified  by
\begin{eqnarray}\label{eqhe}
0\,=\,\nabla \times \nabla\Phi  & = & \nabla\times(\frac{ \nabla\Phi_{N} + \nabla\times \vec{{h}}}{\mu(\frac{|\nabla\Phi|}{a_0})})\,.
\end{eqnarray}
In the two-body approximation to the Earth-Moon system, considering the fact that the mass of Moon is much smaller than the Earth's mass, and higher gravitational moments can be neglected, the Newtonian gravitational field around the earth is spherical. This in turn implies that  $\vec{h}$ is vanishing for the Earth-Moon system. So the gravitational acceleration near the earth, in the AQUAL model holds
\begin{eqnarray}\label{EMond1}
\mu(\frac{|\nabla\Phi|}{a_0}) \nabla\Phi  & = & \nabla\Phi_{N} \,.
\end{eqnarray}
Since near the Earth it holds $\mu\approx 1$ then \eqref{EMond1} can be perturbatively solved for $\nabla \Phi$ 
\begin{eqnarray}\label{EMond2}
 \nabla\Phi  & \approx &\frac{1}{ \mu(\frac{|\nabla\Phi_N|}{a_0})} \nabla\Phi_{N} \,.
\end{eqnarray}
The epsilon expansion in \eqref{epsilonexpansions} provides a sub leading correction to  \eqref{EMond2}. 
Utilizing $\nabla \Phi_N= \hat{r} \frac{GM_\oplus}{r^2}$  results  
\begin{eqnarray}\label{EMond3}
 \nabla\Phi  & \approx &\frac{\hat{r}}{ \mu(\frac{GM_\oplus}{a_0 r^2})}  \frac{GM_\oplus}{r^2}  \,.
\end{eqnarray}
Recalling that $\partial_r \Phi = - \frac{1}{2} \frac{d g_{tt}}{dr}$ and $g_{tt}= c^2 (1-\frac{r_h}{r}+g^{(m)}_{tt})$ then results 
\begin{equation}
g^{(m)(0)}_{tt}(r) - g^{(m)(0)}_{tt}(r_\oplus) \,=\,  (\frac{GM_\oplus a_0}{c^4})^{\frac{1}{2}} \int_{x_\oplus}^{x_r} \frac{dx}{\sqrt{x}}(\frac{1}{\mu(x)}-1)\,,
\end{equation}
where
\begin{equation}
x = \frac{G M_\oplus}{a_0 r^2}\,,
\end{equation}
and $r_\oplus$ is the radius of the Earth. We choose the unites of time and space  such that 
\begin{equation}
g^{(m)(0)}_{tt}(r_\oplus) = 0\,. 
\end{equation}
We are interested to know $g^{(m)(0)}_{tt}(r)$ for $r_\oplus <r < r_{_{LD}}$ where $r_{_{LD}}$ is the lunar distance. In this regime  
\begin{eqnarray}
\label{xld}
x_{_{LD}} &=& \frac{GM_\oplus}{a_0 r_{_{LD}}^2}\,=\,2.7 \times 10^{7},\\
x_{r_\oplus} &=& \frac{GM_\oplus}{a_0 r_\oplus^2}\,=\, 9.6 \times 10^{10},\\
x_{_{LD}}  &<x_r<& x_{r_\oplus} ,
\end{eqnarray}
so we can expand $\mu(x)$ in large $r$ and keep the leading and the first sub leading terms:
\begin{subequations}\label{muchoice}
 \begin{eqnarray}
 \mu_1(x) &\approx & 1 -\frac{1}{2 x} \,,\\
 \mu_2(x) &\approx & 1 -\frac{27}{ x^3} \,,\\
 \mu_\alpha(x) &\approx & 1 -\frac{1}{\alpha x} \,,\\
 \mu_n(x) &\approx & 1 -\frac{1}{n x^n} \,.
 \end{eqnarray}
 \end{subequations}
Eq. \eqref{mux} and eq. \eqref{y=xmux} at very large $x$ implies that  
 \begin{eqnarray}
 y_{r} &\approx& x_{r} ~\to ~ \nu(x_{r})\approx \frac{1}{\mu(x_{r})}\,, 
  \end{eqnarray}
which in turn yields:
\begin{subequations}\label{nuchoice}
\begin{eqnarray}
\label{3.21a}
\tilde{\mu}_n(x) \approx \frac{1}{\nu_n(x)} &=& 1- \frac{1}{n x^n}\,, \\
\tilde{\mu}_\beta(x) \approx \frac{1}{\nu_\beta(x)} &=& 1-(1/2+\beta) e^{-x} \,,\\
\tilde{\mu}_\gamma(x) \approx \frac{1}{\nu_\gamma(x)} &=& 1- e^{- x^{\frac{\gamma}{2}}} \,, \\
\tilde{\mu}_\delta(x) \approx \frac{1}{\nu_\delta(x)} &=& 1 - \frac{1}{\delta} e^{- x^{\frac{\delta}{2}}}\,.
\end{eqnarray}
\end{subequations}
For all the realistic range of the parameters in the choices of \eqref{muchoice} and \eqref{nuchoice}, one finds that 
\begin{equation}
\int_{r_\oplus}^{x_r} \frac{dx}{\sqrt{x}} (\frac{1}{\mu(x)}-1) \ll 1\,.
\end{equation}
This implies that 
\begin{equation}
|g^{(m)(0)}_{tt}(r)| \,\ll \,  (\frac{GM_\oplus a_0}{c^4})^{\frac{1}{2}} ~ \to~ |g^{(m)(0)}_{tt}(r)| \ll 2.21 \times 10^{-15}\,.
\end{equation}
We assume that 
\begin{equation}
|g^{(m)(0)}_{rr}(r)| \approx |g^{(m)(0)}_{tt}(r)| \ll 2.21 \times 10^{-15}\,.
\end{equation}
We notice that  $g^{(m)(0)}_{rr}(r)$ and $g^{(m)(0)}_{tt}(r)$ contribute to measuring the distance of the Earth-Moon at the order of
\begin{equation}
\Delta L = r_{_{LD}} |g^{(m)(0)}_{rr}(r)| \ll 10^{-6} \mbox{meters}\,,
\end{equation}
wherein the lunar distance (LD) is set to $384,400$ kilometers. 
Notice that adding the $\epsilon$ expansion \eqref{epsilonexpansions} will not significantly change $\Delta L$.  The LLR and LAGEOS satellites identify the center of the mass of the Earth with the precision of centimeters.  Therefore, the deviation of the geometry from that predicted from Einstein general relativity can be neglected.

The gravitational field of the Earth, the Sun, the Moon and other planets can cancel each other in a spacial window of order few centimeters.  These windows are called the MOND windows. Ref. \cite{Galianni:2011ch}  reports that the size of MOND window at the gravitational saddle points of the Moon and the Earth is about few centimeters. However, identifying the  exact time and place of the Earth-Moon system requires taking into account the higher gravitational moments of the Earth and Moon \cite{Earth,moon}.  The light ray that is used to measure the Moon's distance  may occasionally pass through this window. The gravitational field strength inside this window is about and less than $a_0$, and its length is about $10 cm$. From a simple dimensional analysis one can estimate  the change of the metric due to the MOND theory inside the MOND to $p_{MOND}=\frac{1}{c^2}a_0*10cm=10^{-28}$. This change occurs in about $10cm$. So it induces about $10^{-29} m$ change in the path of light.  The current precision of measuring the Earth-Moon distance is many orders of magnitude larger than $p_{MOND}$.\footnote{Observing pulsars through the gravitational saddle point (MOND's window) of the Sun-Jupiter system -whose size is about 100 km- has a better chance to empirically constrain the interaction of light with the  physics of the MOND windows.} The LLR and LAGEOS data we will use in section \eqref{constraints}, are also averaged over a long period of time.  Therefore occasional passage through the MOND window will not affect them. 

\section{The Lunar System Constraint}
\label{constraints}
In the AQUAL theory it is the eq. \eqref{EMond2} that represents the gravitational field around the Earth. 
Now note that \eqref{GMLLR}  represents $r^2\nabla\Phi$ at the Lunar distance:
\begin{eqnarray}\label{1}
\frac{GM_\oplus}{\mu(\frac{GM_\oplus}{a_0 r_{_{LD}}^2})}= 398600.443\pm 0.004 \frac{km^3}{s^2}\,,
\end{eqnarray}
 while  \eqref{GMLAGEOS} represents $r^2\nabla\Phi$ at $r=2 r_\oplus$:
 \begin{eqnarray}\label{2}
\frac{GM_\oplus}{\mu(\frac{GM_\oplus}{4 a_0 r_\oplus^2})}=  398600.4419 \pm 0.0002 \frac{km^3}{s^2}\,,
\end{eqnarray}
Consistency between \eqref{1} and \eqref{2} then demands that 
\begin{eqnarray}\label{LunarCons}
|\frac{1}{\mu(x_{_{LD}})} -  \frac{1}{\mu(x_{2r_\oplus})} | < 10^{-8}\,,
\end{eqnarray}
where $x_{_{LD}}$ is given in \eqref{xld} and
\begin{eqnarray}
%x_{_{LD}} &=& \frac{GM_\oplus}{a_0 r_{_{LD}}^2}\,=\,2.7 \times 10^{7}\\
x_{2r_\oplus} &=& \frac{GM_\oplus}{4 a_0 r_\oplus^2}\,=\, 2.4 \times 10^{10}\,,
\end{eqnarray}
wherein   the 
radius of the earth is taken to be $6,371$ kilometers.

Eq. \eqref{LunarCons} is the Lunar system constraints on the AQUAL functional. $\mu_1$ and $\mu_2$ hold
\begin{eqnarray}
|\frac{1}{\mu_1(x_{_{LD}})} -  \frac{1}{\mu_1(x_{2r_\oplus})} | & = & 1.85 \times 10^{-8} \,,\\
|\frac{1}{\mu_2(x_{_{LD}})} -  \frac{1}{\mu_2(x_{2r_\oplus})} | & = & 1. 37 \times 10^{-21} \,.
\end{eqnarray}
So the choice of $\mu_1$ is refuted at the confidence level of 1.85 sigmas while $\mu_2$ remains consistent with the Lunar system constraints. Eq. \eqref{LunarCons} also constraints $\alpha$ and $n$ in \eqref{alphafamily} and \eqref{nfamily}:
 \begin{eqnarray}
 %|\frac{1}{\mu_\alpha(x_{_{LD}})} -  \frac{1}{\mu_\alpha(x_{2r_\oplus})} | < 10^{-8} &\to&
   \alpha > 3.69954\,, \\
 %|\frac{1}{\mu_n(x_{_{LD}})} -  \frac{1}{\mu_n(x_{2r_\oplus})} | < 10^{-8} &\to& 
  n > 1.07237 \,.
 \end{eqnarray}
 Due to \eqref{3.21a}, \eqref{LunarCons}  implies $n>1.072$  for  \eqref{nun} . Since $\nu_\beta(y)-1$, $\nu_\gamma(y)-1$ and $\nu_\delta(y)-1$ are exponentially suppressed   in large y, \eqref{LunarCons} provides no strong  constraint on them. Those results, however, can be interpreted as  limits on the  intermediate-range gravity  \cite{Li:2005tq}. Note that  \cite{Li:2005tq} has used the same set of data but has applied it to a single choice for the interpolating function.  Ref. \cite{Famaey:2011kh} within its 6.4 section reviews the Solar constraints on $\mu$. Though the Lunar system constraint presented here is weaker than the constraint due to the perihelion precession of the Saturn reported in  \cite{Blanchet:2011pv,Fienga:2009ub}\footnote{They constrain \eqref{nfamily} to $n>8$.}, note that here we have used  a different set of data. Furthermore our constraint is directly derived from the MOND effect, and in contrary to \cite{Blanchet:2011pv,Fienga:2009ub} is not based on to the external field effects.

Let us apply the Lunar system constraint on the standard, the standard and the Bekenstein's choices \cite{Zhao:2005gk}:
\begin{eqnarray}
\mu_{\text{standard}} &=& \mu_n(x)|_{n=2}= \frac{x}{\sqrt{1+x^2}}\,, \\
\mu_{\text{simple}} & =& \mu_n(x)|_{n=1}=  \frac{x}{1+x}\,,\\
\mu_{\text{Bekenstein}} &=& \mu_{\alpha}(x)|_{\alpha=0}  = \frac{\sqrt{1+4x}-1}{\sqrt{1+4x}+1}\,.
\end{eqnarray}
  The Lunar system constraint is consistent with the standard choice. It, however, refutes the simple choice with the  confidence level of $3.69$ sigmas. It refutes the Bekenstein choice at the confidence level of    18600 sigmas.

\section{Conclusions}
We have translated the GM mass measured by the  Lunar Laser ranging measurements and LAGEOS satellites into a  constraint on the interpolating function of modified gravities. The Lunar constraint is given in \eqref{LunarCons}. We have applied the Lunar system's constraint on all the families of the interpolating function  considered in the literature and reviewed in \cite{Famaey:2011kh}. We have proven that the Lunar system constraint is non-trivial for a set of families.  In particular we have proven that the Lunar system constraint refutes the simple interpolating function at the confidence level of   the  $3.69$ sigmas, and completely refutes the Bekenstein choice.

\acknowledgments
This work was supported by the Institute for Research in Fundamental Sciences (IPM).

% The bibliography will probably be heavily edited during typesetting.
% We'll parse it and, using the arxiv number or the journal data, will
% query inspire, trying to verify the data (this will probalby spot
% eventual typos) and retrive the document DOI and eventual errata.
% We however suggest to always provide author, title and journal data:
% in short all the informations that clearly identify a document.

\end{document}